# Protein-protein modelling using cryo-EM restraints

"Cryo-EM driven protein docking"

Mikael Trellet, Gydo van Zundert and Alexandre M.J.J. Bonvin*

Computational Structural Biology Group, Bijvoet Center for Biomolecular Research, Faculty of Science - Chemistry, Utrecht University, Padualaan 8, 3584 CH Utrecht, the Netherlands.

\* Phone: +31.30.2533859, Fax: +31.30.2537623, Email: a.m.j.j.bonvin@uu.nl

## i. Summary

The recent improvements in cryo-electron microscopy (cryo-EM) in the past few years are now allowing to observe molecular complexes at atomic resolution. As a consequence, numerous structures derived from cryo-EM are now available in the Protein Data Bank. However, if for some complexes atomic resolution is reached, this is not true for all. This is also the case in cryo-electron tomography where the achievable resolution is still limited. Furthermore the resolution in a cryo-EM map is not a constant, with often outer regions being of lower resolution, possibly linked to conformational variability. Although those low to medium resolution EM maps (or regions thereof) cannot directly provide atomic structure of large molecular complexes, they provide valuable information to model the individual components and their assembly into them. Most approaches for this kind of modelling are performing rigid fitting of the individual components into the EM density map. While this would appear an obvious option, they ignore key aspects of molecular recognition, the energetics and flexibility of the interfaces. Moreover, these often restricts the modelling to a unique source of data, the EM density map.

In this chapter, we describe a protocol where an EM map is used as restraint in HADDOCK to guide the modelling process. In the first step rigid body fitting is performed with PowerFit in order to identify the most likely locations of the molecules into the map. These are then used as centroids to which distance restraints are defined from the center of mass of the components of the complex for the initial rigid-body docking. The EM density is then directly used as an additional restraint energy term, which can be combined with all the other type of data supported by HADDOCK. This protocol relies on the new version 2.4 of both the HADDOCK webserver and software. Preparation steps consisting of cropping the EM map and rigid-body fitting of the atomic structure are explained. Then, the EM-driven docking protocol using HADDOCK is illustrated.

# 1. Introduction.

To drive all essential functions of the cells, biomolecules interact with each other forming complexes of different scales and stabilities. Deciphering the three dimensional (3D) structure of such molecular complexes provides insights into the molecular determinants of these interactions and opens the route to tuning them in order to prevent or promote functions linked for example to diseases. Several experimental techniques exist to solve the 3D structure of molecules. Depending on the flexibility, mobility and environment of those proteins, some techniques will be more efficient than others. They might also picture the system at different resolutions. X-ray crystallography and NMR have been for a long time the sole providers of high-resolution atomic structures stored in the Protein Data Bank (PDB). However, the past few years have seen the rise in the number of high-resolution structures solved by cryo-electron microscopy (cryo-EM). Cryo-EM has undergone a revolution in terms of the achievable resolution thanks to both technical (e.g. the direct electron detectors) and software advances (Bai, McMullan, and Scheres 2015; Kimanius et al. 2016).

Despite those advances, there will still be plenty of cases where cryo-EM will not achieve atomistic resolution (also typically difficult to reach in cryo-electron tomography). The resolution within one large macromolecular complex is also not a constant, meaning that parts of the complexes, often on the periphery or the more flexible parts, might only be seen at lower resolution. In those cases one has to rely to fitting structures or models of the components of a complex into the density. This can be done via different ways: Manual fitting using specialized tools (Pettersen et al. 2004; Baker and Johnson 1996), exhaustive search and rigid-body fitting (Esquivel-Rodríguez and Kihara 2013) or flexible fitting, using different strategies to account for the atomic structures flexibility (McGreevy et al. 2016). Often this modelling does not take into account flexibility (or only to a limited degree) and usually ignores the energetics at the interface of the fitted components, with the result that the interfaces in those complexes often have a poor quality with many clashes.

We have previously published a protocol that makes use of cryo-EM densities in flexible docking based on our information-driven, integrative modelling platform HADDOCK (van Zundert, Melquiond, and Bonvin 2015). In this chapter, we illustrate the use of cryo-EM data as restraints to drive the modelling of a protein-protein complex using the new HADDOCK2.4 web portal, which now support such kind of data. The protocol illustrates various steps, from the preparation/cropping of the original cryo-EM map, rigid-body fitting into the cryo-EM density to extract centroids position, and finally the setup of HADDOCK-EM run using its web portal version.

## 2. Overview

This section describes the different steps and their background in order to perform a protein-protein docking run in HADDOCK using an EM density map as restraint.

HADDOCK makes use of a variety of restraints (often expressed in terms of ambiguous or unambiguous distance restraints) throughout the entire docking process to drive and score the complex formation. These restraints can be derived from various experimental information sources such as NMR chemical shifts perturbations, hydrogen/deuterium exchange, chemical cross-linking detected by mass spectrometry, mutagenesis, etc. (van Dijk, Boelens, and Bonvin 2005; Melquiond and Bonvin 2010; Karaca and Bonvin 2013; Rodrigues and Bonvin 2014).

When using cryo-EM data, however, HADDOCK needs to first convert the information provided by the EM map into distance restraints in order to drive the molecules to their potential location. This can be done by extracting centroids from the EM map as described in (C.P.van Zundert and M.J.J. Bonvin 2015). The centroids are provided as 3D coordinates to HADDOCK, and are automatically converted to unambiguous (or ambiguous in cases where circular symmetry is present or the identity between subunits is uncertain) distance restraints between the centroids and the center of mass of the subunits, as illustrated in **Figure 1**. These restraints draw, during the initial rigid-body step of HADDOCK, the molecules toward their location within the EM map. Once the rigid complex is formed and oriented correctly in the density, the cryo-EM density-based restraint energy term in HADDOCK is applied, and the refinement protocol proceeds through the various steps of HADDOCK. For details see section 2.3.and the original HADDOCK-EM publication (van Zundert, Melquiond, and Bonvin 2015).

### 2.1 High-resolution atomic structures rigid-body fitting into cryo-EM densities

The rigid-body fitting into the cryo-EM map will be performed using PowerFit (C.P.van Zundert and M.J.J. Bonvin 2015), making use of our web server (van Zundert et al. 2017). PowerFit fits atomic structures into density maps by performing a full-exhaustive 6-dimensional cross-correlation search between the atomic structure and the density. It takes as input an atomic structure in PDB- or mmCIF-format and a cryo-EM density with its resolution, and outputs positions and rotations of the atomic structure corresponding to high correlation values and the top 10 best scoring rigid poses. PowerFit uses the local cross-correlation function as its base score. The score is by default enhanced with an optional Laplace pre-filter, and a core-weighted version that minimizes the effect overlapping densities from neighboring subunits.

From the fitted structure one can extract the 3D coordinates of the centroids (their center of mass position into the map), an information required by HADDOCK-EM.

## 2.2 Cryo-EM density map cropping

In order to reduce data noise and save computational time, we strongly advise to crop the cryo-EM map to the region of interest. Cropping can be straightforwardly performed using UCSF Chimera (Pettersen et al. 2004). A step-by-step protocol to extract a subregion of a density map is available at https://www.cgl.ucsf.edu/chimera/docs/UsersGuide/midas/mask.html. In this protocol we will use fitting results from PowerFit to crop the map with respect to the predicted molecular subunits" location.

## 2.3 Protein-Protein HADDOCKing with EM restraints

### 2.3.1 Docking protocol

The docking protocol in HADDOCK consists of three successive steps:
- *it0*: Rigid-body energy minimization (RBEM)
- *it1*: Semi-flexible simulated annealing (SA) in torsion angle space (TAD/SA)
- *water*: Final restrained molecular dynamics in explicit solvent

Pre- and post-processing steps are performed: *i)* to build missing atoms in the preliminary step and *ii)* to launch a variety of analyses and clustering of solutions in the final step. For further details please refer to (Dominguez, Boelens, and Bonvin 2003; S. J. de Vries et al. 2007).

The HADDOCK-EM protocol requires as input an EM density map and its resolution together with the centroid coordinates of each of the subunits to be docked. Some changes have been made to the default HADDOCK docking protocol to account for the cryo-EM data parameters, mainly in it0, where centroids, approximate location of the subunits COMs in the density map obtained during the fitting step (see Section 2.2) are used to place the subunits. As for the center of mass docking protocol of HADDOCK (Sjoerd J. de Vries et al. 2010), additional distance restraints are generated between the COMs of the subunits. The main difference here lies in the fact that distances restraints are not created between the subunits themselves but between each subunit and one or several (in case of ambiguity) centroid coordinates.

Other cryo-EM-related required parameters for HADDOCK are either directly extracted from the map or have optimised default values. Some of these can be controlled through the web portal interface, for expert tuning of results.

*2.3.1.1 Rigid-Body Energy Minimization (RBEM, it0)*

In the initial docking stage, the interacting partners are considered rigid and separated in space and placed on a sphere centred on the midpoint of the centroids. For each docking trial, each subunit is randomly rotated around its center of mass and translated within a 10Å box of to ensure unbiased starting configurations. In the case of unambiguous centroid-based restraints, HADDOCK will fit the subunits' COMs on the centroids to which they are associated. In the case of ambiguous restraints each subunit would be ambiguously linked to any of the centroid given as input. Then, selection of the best conformation will solely rely on the HADDOCK score.

The centroid-based distance restraint is described by a soft square potential between two pseudo-atoms, one of which corresponds to the centroid and the other to the COM of the subunit.

Optimisation steps have been performed to derive the best values for (1) the force constant of the centroid-based distance restraints that drives the COMs to the centroids, (2) the weight for the cross-correlation energy term and (3) the weight of the LCC term in the HADDOCK score for *it0*. The default values in our protocol stands respectively at 50, 15000 and 100. Those 3 values can be changed in the submission interface of HADDOCK2.4.

Binary systems will undergo a supplementary optimization step that aims at optimizing their orientation within the EM map. For this, an exhaustive 4 degrees rotation search along the axis joining the centroids is performed and at each step the cross-correlation value is calculated to assess the pose. The orientation with the maximal cross correlation value is kept. Finally, a rigid-body minimisation is performed against the map using a combination of the cross-correlation, van der Waals and electrostatic energy terms. Models are then scored by the traditional HADDOCK score plus a LCC term that reports on the overall quality of the fitting within the EM map. Typically, 2,000 models are generated and scored from which typically the 400 models with the best HADDOCK score (see section 2.3.1.4) will go to the semi-flexible simulated annealing stage of HADDOCK.

*2.3.1.2 Semi-Flexible Simulated Annealing in Torsion Angle Space (TAD/SA, it1)*

After a first rigid-body simulated annealing stage, the semi-flexible simulated annealing stage, which starts with a short rigid-body molecular dynamics phase, optimizes the side chain conformations at the interface and then both backbone and side-chains. The flexible regions are automatically defined for each docking model as the residues within 5Å from a partner molecule. The parameters for *it1* are the same as in a typical docking run with HADDOCK, with the exception of adding the cross-correlation energy term used both during the simulated annealing protocol and in the scoring.

*2.3.1.3 Restrained Molecular Dynamics in Explicit Solvent (water)*

The structures obtained after simulated annealing are finally refined in an explicit solvent layer to further improve the scoring. This is done by a short molecular dynamics simulation in water, solvating the complex in an 8Å shell of TIP3P water molecules (Jorgensen et al. 1983).

*2.3.1.4 Scoring*

The EM protocol introduces a new term to the HADDOCK score, namely the local cross-correlation value (LCC) computed for a given model which is added to the equation defining the score, with an optimal weight for the three stages:

$$HS_{EM-it0} = 0.01 * E_{vdw} + 1.0 * E_{elec} + 0.01 * E_{AIR} + 1.0 * E_{desolv}$$
$$- 0.01 * BSA - 400 * LCC$$
$$HS_{EM-it1} = 1.0 * E_{vdw} + 1.0 * E_{elec} + 0.1 * E_{AIR} + 1.0 * E_{desolv}$$
$$- 0.01 * BSA - 10000 * LCC$$
$$HS_{EM-itw} = 1.0 * E_{vdw} + 0.2 * E_{elec} + 0.1 * E_{AIR} + 1.0 * E_{desolv}$$
$$- 10000 * LCC$$

The other terms of the scoring function are the intermolecular van der Waals ($E_{vdw}$) and electrostatic ($E_{elec}$) energies calculated with the OPLS force field and an 8.5Å non-bonded cutoff (Jorgensen and Tirado-Rives 1988), an empirical desolvation potential ($E_{desolv}$) (Fernández-Recio, Totrov, and Abagyan 2004), the ambiguous interaction restraints energy ($E_{AIR}$), and the buried surface area (BSA).

### 2.3.2 Clustering of Final Solutions

All models generated by HADDOCK are clustered either based on their fraction of common contacts (Rodrigues et al. 2012) (FCC, default) or on their interface-ligand-RMSD (i-l-RMSD) depending on the user's choice.

## 3. Methods

The HADDOCK-EM protocol requires some preliminary steps outside the traditional HADDOCK pipeline and independent from the web server. As explained in the previous sections, atomic structures will first be fit into an EM map region, then the EM map will be cropped, followed by a final fitting step.

To follow our protocol in its entirety, the 3D viewer program UCSF Chimera (https://www.cgl.ucsf.edu/chimera/) is needed. The protocol described here is based on version 1.12.0. Python2 should also be installed (we recommend the latest stable version Python 2.7.15). All other steps will simply make use of a standard web browser with JavaScript enabled. A registration to the CSB portal is required to use both PowerFit and HADDOCK (see **Note 1**). Complementary to the HADDOCK registration, users must request GURU access via their profile page to get access to the EM restraints parameters.

In the following sections, we illustrate our protocol on a test case taken from the use cases illustrated in (van Zundert, Melquiond, and Bonvin 2015). The complex studied describes the interaction between two proteins of the 30S subunit of the ribosome (chains F and R). An atomic model of the entire complex is available (PDBid: 2YKR) as well as the 9.8Å resolution cryo-EM map from which it was derived (EMDBid: 1884). The necessary files are provided in a tar archive available in the Supplementary Material. The protocol should be able to run on any operating system.

**3.1 Pre-processing of the cryo-EM map**

In this section, we will crop the cryo-EM map to only keep the part that is relevant for our docking. This step is optional and significantly depends on 1) the size of the map and 2) the preliminary information we have about the structures localisation within this map. In our example, we already know the location of the subunits we want to dock in the EM map. Without this information, a very first step would have been to perform a fitting of the subunits within the EM map we plan to use to identify their possible location. Such fitting should always start from the largest components since these are easier to identify in the EM map.

To crop the map, we will use UCSF Chimera. Chimera has a very complete support for density maps and allows to quickly observe, analyse and manipulate such maps via a customised user interface. We will follow the instructions given in the Chimera documentation (Goddard, Huang, and Ferrin 2005) with little modifications accounting for most recent versions of Chimera. For an online version of the documentation please refer to https://www.cgl.ucsf.edu/chimera/docs/UsersGuide/midas/mask.html.

1. Open Chimera
2. Load the cryo-EM map (`emd_1884.map`)
3. Load the crystal structure of the 30S subunit bound to RsgA (`2ykr_FR.pdb`).

> *WARNING: At this stage, be careful to not move the complex independently from the EM map during the session. This could lead to erroneous results during the next steps.*

4. If the *Model Panel* window is not displayed, go to *Tools > General Controls > Model Panel.*
5. Select "2ykr_FR.pdb"
6. Click on *Action > Surface > Show* to generate the surface of the protein.
7. A new line should appear in the *Model Panel* window with the name "MSMS main surface of 2ykr_FR.pdb"
8. Click on *Tools > General Controls > Command Line.*
9. In the new dialog window that opened at the bottom of the viewer main window type:

    `mask #0 #1`

    where `#0` represents the identifier of your EM map and `#1` the identifier of your protein/protein surface.
10. A new volume representation should appear (see **Figure 2**) and together with a new line in the *Model Panel* window named "emd_1884.map masked"
11. Save the new masked map, in the *Volume Viewer* window: *File > Save map as...* (`1884_masked.map`).

## 3.2 Getting centroids coordinates by fitting the atomic structure into the new cryo-EM map

In this section we will use the PowerFit web server to obtain the centroids coordinates of the two subunits of the complex. PowerFit performs an exhaustive search to identify the best fit of our crystal structure within the new masked cryo-EM map obtained in section 3.1. The best solutions are ranked according to a cross-correlation score (similar to the one used in the HADDOCK protocol).

Note that access to the PowerFit web server requires registration (https://nestor.science.uu.nl/auth/register/ - select **PowerFit** as registered service).

1. Go to
    http://milou.science.uu.nl/cgi/services/POWERFIT/powerfit/submit

2. Add the cryo-EM map file (`1884_masked.map`) to the *"Cryo-EM map"* field.
3. Add the atomic structure of the complex (`2ykr_F.pdb`) to the *"Atomic structure"* field.
4. Put `9.8` as *"Map resolution"* (in Angstroms).
5. By default the server will redirect the computation to GPGPU grid resources provided by the federated sites of EGI. To run locally on our server you might choose to uncheck *"Redirect submission to grid (GPU) resources"*
6. Enter your credentials (email + password) and click on *"Submit"*.
7. The run should take about 5 minutes. The status of your job will be updated every 30 seconds. Once the job is finished, you will get an email and, if you have left the page open, you will be redirected to the results page, similar to the one shown in **Figure 3**.
8. On this page, reach the *"Solutions"* section. The table presented here reports the 15 best non-redundant solutions ranked by correlation score. We will focus on the best solution.
9. Click on the first link of the page corresponding to *"Archive of the complete run"*. This will download the output of PowerFit under an archive file.
10. Untar the archive
11. OPTIONAL: Open files `lcc.mrc` and `fit_1.pdb` with Chimera. Check that the atomic structure is well fitted within the density map file.
12. Redo steps 1 to 10 by only changing the PDB file provided in step 3. But give this time the other protein, `2ykr_R.pdb`, as atomic structure input.
13. Using a terminal (or the windows command-prompt) run the python script "`em_tools/centroid-from-structure.py`" providing the best fit chains PDB files (`fit_1.pdb`) in each run archive as unique arguments.

```
> python centroid-from-structure.py fit_1.pdb
Parsed file:  fit_1.pdb
Corresponding centroid (x, y, z):
11.90   -2.48   75.54
> python centroid-from-structure.py fit_1.pdb
```

```
Parsed file:  fit_1.pdb
Corresponding centroid (x, y, z):
17.40   5.80    58.00
```

14. Save the centroid coordinates for later.

### 3.3 Preparation of input files

Each PDB provided to HADDOCK has to respect the PDB format with proper syntax and clear chain identifiers (*see* **Note 1**). The two input chains for the docking run are the chains F and R of 2YKR and are respectively provided in files `2ykr_F.pdb` and `2ykr_R.pdb`.

The PDB file of the protein must be checked to avoid any double occupancies or residue insertion codes. If present, these can be removed by manual editing of the file, or automatically by using the `pdb_delocc.py` script provided as part of the PDB-tools repository maintained by the HADDOCK team (https://github.com/haddocking/pdb-tools).

The EM map obtained after the previous step of cropping can be submitted as it is. The HADDOCK2.4 new web server processes and converts automatically any map under MRC or CCP4 format to XPLOR format, the latter being the only one read by CNS (Crystallography and NMR System)(Brünger et al. 1998), the computational engine used by HADDOCK.

### 3.4 Docking two subunits of the 30S ribosome with the HADDOCK2.4 web server

For this docking, we will make use of the new HADDOCK web server available in its beta version at (https://haddock.science.uu.nl/services/HADDOCK2.4/). Registration is required to make use of the new interface and can be accessed through the corresponding submenu in the portal. Following the activation by the HADDOCK support team, users must request GURU access to be able to use EM restraints. This can be done in their own user profile page.

1. Open an Internet browser and go to https://haddock.science.uu.nl/services/HADDOCK2.4/. Click on the Submit subsection. You will find the page illustrated in **Figure 4**.

2. We advise to give a name to your docking run. Be aware that no space or special characters other than "-" or "_" are allowed. We propose here to name the run "`2ykr_em_modelling`".
3. There is no precise order for the molecule, either of the PDB file can be provided first, but we do advise as a general rule to provide the largest component as first molecule (see **Note 1)**. By default, we will use chain F as 1st molecule. In the section *"First molecule",* at the entry *"Where is the structure provided?"* Leave option `I am submitting it`. Leave *"Which chain of the structure must be used?"* to `All` (see **Note 2)**. Next to *"PDB structure to submit"* press the `Choose file` button and move to the location where the tutorial data were unpacked. Go to the *pdbs/* directory and select the *2ykr_F.pdb* file. Keep both Nter and Cter to `False`.
4. In the section *"Second molecule",* at the entry *"Where is the structure provided?"* Leave option `I am submitting it`. Leave *"Which chain of the structure must be used?"* to `All` (see **Note 2)**. Next to *"PDB structure to submit"* press the *"Choose file"* button and move to the location where the tutorial data were unpacked. Go to the *pdbs/* directory and select the *2ykr_R.pdb* file. Keep both Nter and Cter to `False`.
5. Click *"Next"* and wait for the second step interface to load (should not take more than a few seconds).
6. Leave the Molecule 1 and 2 parameters empty. Go to section *"EM restraints (optional)"* and unfold it as illustrated in **Figure 5**.
7. Check `Use density/XREF restraints?`
8. Next to *"EM map"* press the *"Choose file"* button and move to the location where the tutorial data were unpacked. Go to the *em_maps/* directory and select the `1884_masked.map` file. (Or select the one you generated at Section 3.2).
9. Set **9.8** in *"Resolution of data in angstrom"* field.
10. If this is not the case, check Use centroid restraints? (set to True)
11. In *"MOLECULE 1 > Centroid position in absolute coordinates"*, enter the coordinates you saved from Section 3.3 for chain A.

12. In "*MOLECULE 2 > Centroid position in absolute coordinates"*, enter the coordinates you saved from Section 3.3 for chain B.
13. Click `Next` and wait for the third step interface to load (should not take more than a few seconds).
14. Leave default parameters and click `Submit` at the bottom of the page.
15. After few seconds you will be redirected to a page reporting the status of your job, a short summary of the docking input and a progression report. This page will be updated every 30 seconds to report the progression of your job.
16. Within typically a few hours, depending on the web server load, you will receive another email reporting the final status of your job. If successful, a result page as depicted in **Figure 6** will be available at the link given in the email or, if you left the status page open, the page will be automatically loaded with a results summary. On this page, you will find the name of your docking run as well as a link to download it as a gzipped tar file. A link to the unique file containing input data and parameters is again provided.
17. The results page also indicates the number of clusters created by HADDOCK and how many structures coming from the *water* steps have been clustered. In our example, **12** clusters are created, gathering **47**% of the top 200 models. For an easier visualization of the results, only the 10 best clusters based on the average HADDOCK score of its top 4 models are displayed in the summary page. You can find information and analyses of the last cluster in the gzipped tar file. For each cluster, information relative to the HADDOCK score of the top 4 models, the cluster size and different statistics and energy values are reported (see **Note 3**).
18. At last, an interactive representation of different CAPRI assessment criteria with respect to the HADDOCK score is provided for the 10 best clusters in the "*Results analysis"* section. An example is shown in the **Figure 6B**. The first three plots show the HADDOCK score versus the Fraction of Common Contacts (FCC – see **Note 4**), the i-RMSD and the l-RMSD calculated using the top ranked model as reference, respectively (see **Note 5**). The last three plots show the van der Waals, electrostatics, and AIRs energy versus i-RMSD. One can note that the Eair values are all equal to 0 because no other restraints than the EM map derived ones have been used to drive this docking.

19. It is possible to manually compare a reference structure with the best models of each cluster generated by HADDOCK. The 3D structures of these models can be directly downloaded from the results page. They are also located in the root of the docking run you downloaded as a gzipped tar file. Their name follows the following syntax: `cluster2_1.pdb`. This file is for instance the best model according to its HADDOCK score in the 2$^{nd}$ cluster given by HADDOCK. The clusters are reported on the result page in the order of their HADDOCK score (from best to worst) (see **Note 6**)

    You can use fitting software such as ProFit (Martin and Porter 2010) to get precise values of RMSD. PyMol is also useful since it has its own fitting algorithm and will give you a RMSD value as well as a visual feedback of the differences between the clustered models and the reference structure. Keep in mind that your reference structure has to be formatted in the same way that the PDB models generated by HADDOCK. ProFit considers only structures with an identical number of atoms.

**Acknowledgments**

This work is supported by European H2020 e-Infrastructure grants (West-Life grant no. 675858 and BioExcel grant no. 675728).

**Notes**

1. Defining the largest molecule as first molecule for docking can be important for the final clustering because, in case of RMSD clustering, the structures are first fitted on the interface residues of the first molecule and then the RMSD is calculated on the interface residues of the second molecule. The interface residues are defined from an analysis of contacts in the generated models (at it1 and water, respectively). Defining the largest molecule first should thus result in a better fitting and clustering. However, one should note that the default clustering method is FCC and the order of the molecules does not impact the FCC calculation algorithm.

2. The PDB files provided to HADDOCK have to be correctly formatted to avoid any issues during the simulation process. There should be no overlap in residue numbering between different chains of a PDB. One can check the proper format of its PDB file using the `pdb_format.py` script provided as part of the PDB-tools repository maintained by the HADDOCK team (https://github.com/haddocking/pdb-tools). Missing atoms in the PDB files are not problematic since HADDOCK will rebuild them automatically.

3. The Z-score indicates how many standard deviations from the average a cluster is located in terms of its HADDOCK score. So the more negative the better.

4. The FCC stands for Fraction of Common Contacts and is calculated by comparing the lists of contacts at the interface between the components of a complex for two different structures. A contact is defined when two residues from different chains of the complex are closer than 5Å from each other. The FCC is calculated as the fraction of common residue pairs shared between the two structures.

5. All reported RMSDs are calculated with respect to the lowest scoring model (the best model according to the HADDOCK score). The i-l-RMSD, which is used for clustering, is calculated on the interface backbone atoms of all chains except the first one after fitting on the backbone atom of the interface of the first molecule. The i-RMSD is calculated by fitting on the backbone atoms of all the residues involved in intermolecular contacts within a cutoff of 10Å. The l-RMSD is obtained by first fitting on the backbone atoms of the first molecule and then calculating the RMSD on the backbone atoms of the remaining chains.

6. The naming of clusters in HADDOCK is linked to their size and not their score. This originates from the clustering software. By definition, the largest cluster is always called cluster1, followed by cluster2 and so on. The cluster size does however not correlate per se with the HADDOCK score. Refer to the result page (or open in a web browser the `index.html` file provided in the tar archive) to see the cluster order based on the HADDOCK score.

**Figure 1:** Representation of the Rigid-Body Docking Protocol in HADDOCK-EM as illustrated in (van Zundert, Melquiond, and Bonvin 2015). (A) Simulated cryo-EM data of colicin E7 / IM7 complex (PDBid 7CEI). (B) Centers of mass of each subunit represented with grey spheres within the EM map. (C) Distance restraints in HADDOCK it0 step are defined between the COM of chain A (light gray) and B (dark gray) and their corresponding centroids. (D) Example of a complex obtained after the first rigid-body minimisation (it0). (E) After the position, the relative orientation of each subunit should be determined. (F) A line drawn between the two centroids is used as axis to perform a rotational search. The complex with the highest cross-correlation value is chosen. (G) Excluding the centroid-based restraints, a final rigid-body minimisation is performed against the cryo-EM data and assessed thanks to a cross-correlation-based potential.

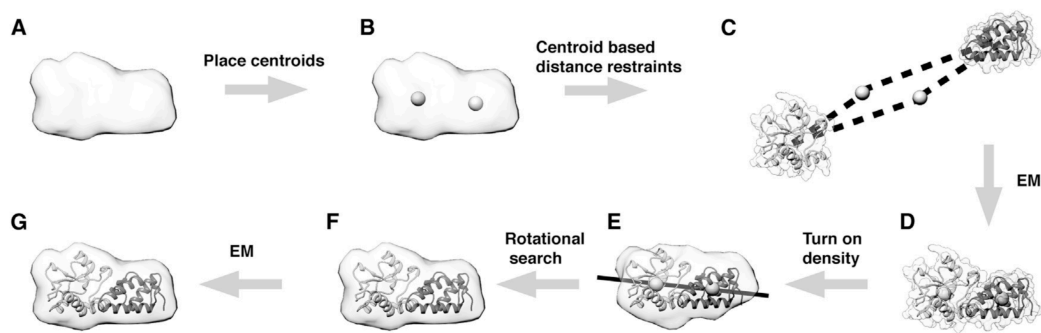

**Figure 2:** Chimera snapshot illustrating the EM map of 30S ribosomal subunit with RsgA bound in the presence of GMPPNP, EMDBid 1884 (Guo et al. 2011), in white and, in blue, subpart of the EMDBid 1884 EM map masked by the subunits F and R of atomic structure 2ykr.

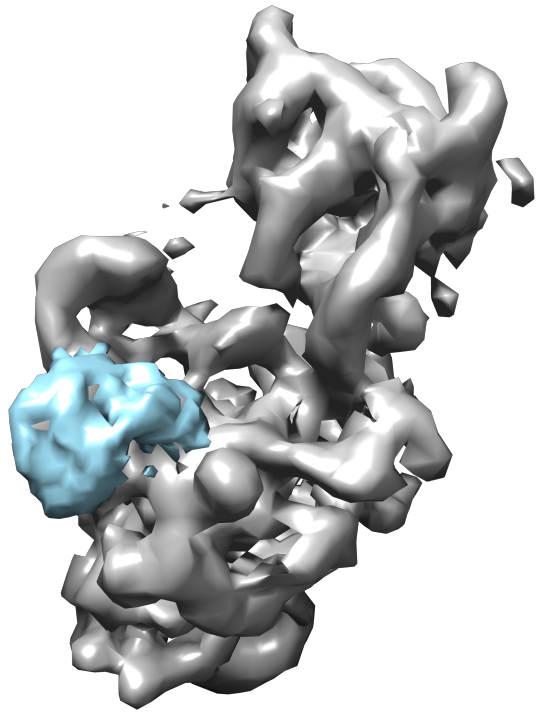

**Figure 3:** Screen capture of PowerFit results page after fitting of chain F of 2ykr in the masked map obtained from EMDBid 1884.

**Figure 4:** Illustration of HADDOCK 2.4 submission page at the *Input data* 1st step.

**Figure 5:** Illustration of HADDOCK 2.4 submission page at the *Input parameters* 2nd step.

**Figure 6:** Illustration of HADDOCK 2.4 results page after docking subunits F and R from 2ykr using as sole restraints the EM map information.

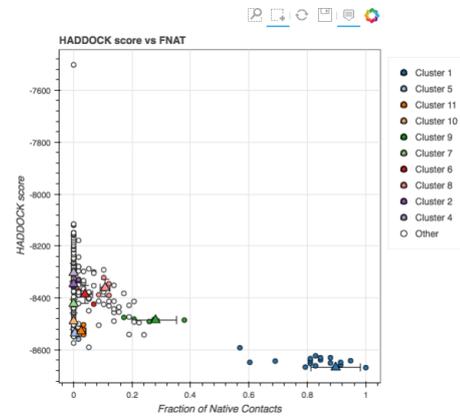

**Figure 7:** Comparison of the best scoring models generated by HADDOCK, in blue (chain F) and green (chain R), and the reference structure (PDBid 2ykr) in dark grey. The EM map used to fit the two subunits and drive the docking run is shown as a transparent surface.

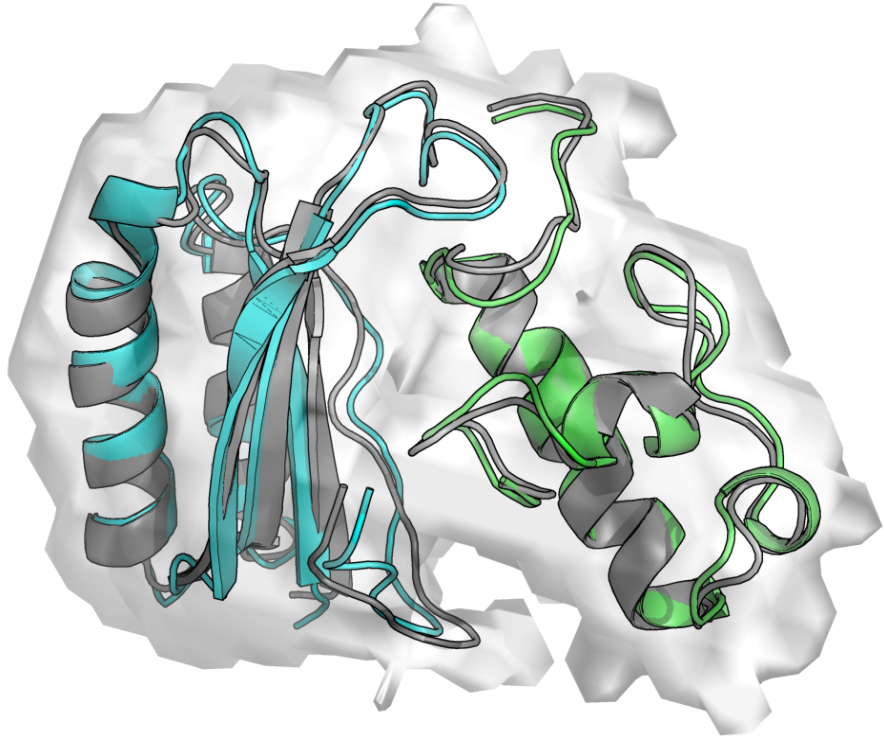